\documentclass[preprint,prb]{revtex4-1}
\usepackage{epsfig}
\usepackage{graphicx}
\usepackage{bm}
\usepackage{amssymb}
\usepackage{amsmath}
\usepackage{amsthm}
\begin{document}
\title{Quenched pinning and collective dislocation dynamics}
\author{Markus Ovaska}
\author{Lasse Laurson} 
\email{lasse.laurson@aalto.fi}
\author{Mikko J. Alava}
\affiliation{COMP Centre of Excellence, Department of Applied Physics,
Aalto University, P.O. Box 11100, 00076 Aalto, Espoo, Finland}
\renewcommand{\abstractname}{\vspace{1.5cm}}
\begin{abstract}
{\bf Several experiments show that crystalline solids deform in a bursty 
and intermittent fashion. Power-law distributed strain bursts in compression experiments of 
micron-sized samples and acoustic emission energies from larger-scale specimens 
are the key signatures of the underlying critical-like collective dislocation 
dynamics - a phenomenon that has also been seen in discrete dislocation dynamics 
(DDD) simulations. Here we show by performing large-scale two-dimensional DDD 
simulations that the character of the dislocation avalanche dynamics changes
upon addition of sufficiently strong randomly distributed quenched pinning centers,
present e.g. in many alloys as immobile solute atoms. For intermediate pinning
strength, our results adhere to the scaling picture of depinning transitions, in
contrast to pure systems where dislocation jamming dominates the avalanche dynamics.
Still stronger disorder quenches the critical behaviour entirely.}
\end{abstract}

\maketitle

The origin of crackling noise \cite{SET-01} in crystal plasticity has an appealing 
explanation in terms of a non-equilibrium phase transition \cite{ZAI-06,ANA-07,ALA-14}: 
If the externally applied stress is high enough, a sample is in a regime of continous 
flow or yielding, while at small stress values and low temperatures one would expect 
the yielding activity to stop after a transient. The existence of a ``yielding 
transition'' separating these two regimes, envisaged to take place at a critical value 
of the applied stress and corresponding to a vanishing plastic deformation rate, would 
then give rise to a natural explanation for the observed scale-free avalanche dynamics 
\cite{UCH-09,DIM-06,CSI-07,WEI-97,MIG-01}. Such a picture has been used in analogy 
to other systems, including the depinning transition of domain walls in disordered 
ferromagnets underlying the magnetic field driven jerky domain walls motion, or 
the Barkhausen effect \cite{DUR-06,LED-02,ROS-09}.

The irreversible deformation process of crystalline solids is a consequence of 
the stress-driven motion of dislocations, line-like defects of the crystal lattice, 
which interact with each other via their anisotropic long-range stress fields. Due 
to these interactions, in combination with constraints due to the underlying 
crystal structure on their motion, dislocations tend to form various complicated 
metastable structures. Thus, the term ``dislocation jamming'' \cite{MIG-02,LAU-10,LIU-01} 
has been coined to describe their tendency to get stuck due to many-body dislocation 
interactions, and this mechanism is then expected to be behind the emergence of a 
finite yield stress in ``pure'' crystals, without a significant population of 
additional defects, such as solute atoms or the complications of e.g. grain boundaries.  
The character of the dislocation jamming transition in such ``pure'' DDD models has 
been analyzed from various angles \cite{MIG-02,LAU-10,LAU-12,ISP-11,ISP-14}.
A recent study found \cite{ISP-14} that the scaling exhibited by 
the strain bursts within a two-dimensional (2$d$) pure DDD model is fundamentally 
different from that expected within the mean field or high-dimensional limit of the 
pinning/depinning scenario \cite{FIS-98}, 
often assumed to describe bursty plastic deformation: 2$d$ dislocation 
dynamics seems to exhibit critical signatures with ``anomalous'' properties not only in the 
proximity of the yielding transition, but also at very low external stresses. A 
detailed study is so far missing in three dimensions, though 3$d$ DDD simulations 
of dislocation avalanches seem to reproduce partially the typical single crystal 
compression results \cite{CSI-07} and to be close to the mean-field case.

\begin{figure}[t!]
\includegraphics[angle=0,width=9.5cm]{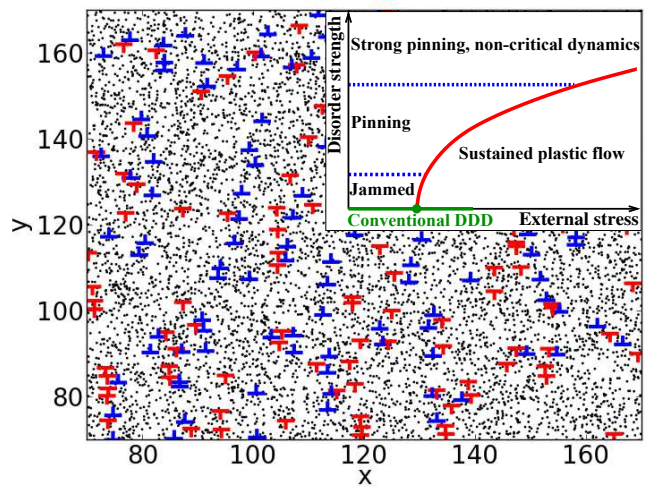}
\caption{\textbf{A snapshot of a two-dimensional dislocation assembly with 
quenched pinning centers, and a phase diagram portraying the nature of dislocation 
dynamics.} The main figure shows a part of the system: Edge dislocation with 
positive and negative Burgers vectors are shown in blue and red, respectively, 
and black dots denote randomly positioned quenched pinning sites (solute atoms). 
The total system size is $L = 200b$ and it contains $N_\text{d} \approx 900$ 
dislocations and $N_\text{s} = 32000$ pinning centers. The inset shows a schematic 
phase diagram in the space spanned by external stress and strength of the quenched 
disorder. As the disorder strength increases for a fixed external stress, jamming 
becomes pinning, and finally critical dynamics ceases at very strong disorder.}
\label{fig:system}
\end{figure}

In reality, plastic deformation or dislocation glide is usually further complicated by 
the presence of various kinds of defects - precipitates, grain boundaries, vacancies, 
solute atoms - that interact with the dislocations and thus interfere with the 
deformation process \cite{ZAP-01,MOR-04}; indeed much of metallurgy is based on the 
practical utilization of this phenomenon to optimize hardness or ductility \cite{COU-90}, 
also known well in the context of high-temperature superconductors with vortex pinning 
\cite{BLA-94}. Here we study the effect of disorder on collective dislocation 
dynamics. To this end, we generalize the standard, two-dimensional (2$d$) DDD 
models \cite{GIE-95,MIG-08} to include a random arrangement of $N_\text{s}$ quenched 
pinning centers. This random pinning landscape could be due to e.g. solute 
atoms with a low mobility; however, the detailed nature of the pinning centers is 
irrelevant for our conclusions and in general one expects on the basis of the theory 
of depinning of elastic manifolds in random media that the microscopic details are 
not important \cite{LED-02,FIS-98}. A snapshot of the 2$d$ system with pinning centers 
interacting with the dislocations \cite{LEY-10} is shown in Fig. \ref{fig:system}. 

The disorder immediately widens the phase diagram of the DDD models at a constant applied
stress so that the (candidate) order parameter, shear rate, becomes a function of both 
the applied shear stress and the relative strengths of the pinning and the 
dislocation-dislocation interactions. For strong enough disorder, we find a yielding 
transition that agrees with a standard depinning scaling scenario, exhibiting the 
typical signatures of power-law distributions for both avalanche 
sizes and durations, with the cut-offs of the distributions displaying power-law 
divergences at a critical applied stress: yielding becomes depinning, and jamming 
becomes pinning. The set of critical exponents characterizing these distributions is 
found to be different from that of mean field depinning. With still stronger disorder, 
the collective behaviour disappears, as is indicated by the qualitative phase diagram 
in the inset of Fig. \ref{fig:system}. A special point of the phase diagram 
corresponding to the low-disorder limit is given by the 
corresponding pure system \cite{ISP-14}, where the scaling behaviour is completely 
different from those of the above-mentioned two phases, where pinning plays a role. 
We next explore these two novel phases, in particular by considering the scaling 
properties of dislocation avalanches.

\begin{figure}[!t]
\includegraphics[angle=0,width=9.5cm,clip]{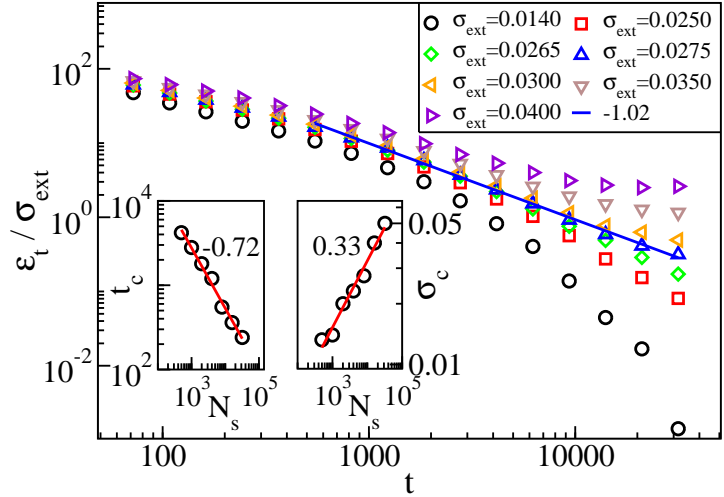}
\caption{\textbf{Relaxation of the order parameter exhibits typical power-law behavior of 
depinning transitions.} Main figure shows the time dependence of the strain rate $\epsilon_t$
for different applied stress values, in a system with $N_\text{s} = 8000$ pinning sites or
solute atoms, of strength $A = 0.05$. For $\sigma_\text{ext}$ close to $\sigma_\text{c} = 0.0275$,
$\epsilon_t$ decays as a power law of time, $\epsilon_t \propto t^{-\theta}$, with
$\theta \approx 1.0$. The insets show the dependence of the initial transient time $t_\text{c}$
after which the $t^{-1}$ power law decay sets in (left) and the corresponding critical
stress $\sigma_\text{c}$ (right) on the number of solutes $N_\text{s}$, while $A$ is kept 
constant.}
\label{fig:2}
\end{figure}

\vspace{0.5cm}
{\bf \large \noindent Results}

\noindent
In order to address and characterize the various aspects of collective dislocation 
dynamics under the influence of disorder, we consider an extension of the standard 
2$d$ DDD model \cite{GIE-95,MIG-08} with single-slip geometry, with the additional 
ingredient of a quenched pinning field, see Fig. \ref{fig:system} and Methods for 
details. The model consists of $N_\text{d}$ edge dislocations with an equal number 
of positive and negative Burgers vectors (with the blue and red symbols in Fig. 
\ref{fig:system} corresponding to the signs of the Burgers vectors, $s_n = +1$ and 
-1, respectively), interacting via their long-range anisotropic stress fields, and 
gliding along the $x$ direction within a square simulation box of linear size $L$. 
The quenched pinning field is modelled by including $N_\text{s}$ randomly distributed 
immobile pinning centers or solute atoms (black dots in Fig. \ref{fig:system}), 
interacting with the dislocations with an interaction strength $A$. Overdamped
dynamics is assumed, such that the velocity $v_n$ of the $n$th dislocation is
proportional to the total stress (with contributions from interactions with other 
dislocations, pinning centers and the external stress) acting on it. 
While not including all the details of full three-dimensional systems with flexible 
dislocation lines \cite{CSI-07}, the model is expected to capture the essential 
features of the crossover from jamming (dislocations getting stuck to each other) 
to pinning (dislocations getting stuck to quenched pinning centers), and is simple 
enough to allow collecting high quality statistics in numerical simulations, an 
advantage of the 2$d$ system over the 3$d$ ones. We have verified that our results 
presented below are robust with respect to changes in details of the pinning 
potential - as expected on the account of the analogy with depinning models - and 
are free of any clear finite-size effects detrimental to our results. Following the 
standard procedure of DDD simulations \cite{MIG-02}, initially random arrangements 
of dislocations are first let to relax in zero external stress, $\sigma_\text{ext}=0$, 
to reach metastable arrangements. Then, the external stress is swithed on, and the 
time-evolution of the system is monitored.\\

{\bf \noindent Depinning transition of the dislocation ensemble.}
Dislocation dynamics in a disordered background is expected to be
complicated, with transients and relaxations typical of glassy systems as is the case
also for the depinning of elastic manifolds. The first issue we explore is the 
response of the system to a constant external stress $\sigma_\text{ext}$ at zero 
temperature ("constant control parameter"). In the steady state, the number of 
dislocations is $N_\text{d} \approx 800 - 900$ within a rectangular system of 
linear size $L=200b$ (with $b$ the magnitude of the Burgers vector of the dislocations). 
To tune the pinning strength, we vary the number of pinning centers/solutes in the 
range $N_\text{s} = 500-32000$ and set $A = 0.05$. Fig. \ref{fig:2} shows that the 
time-dependent strain rate $d\epsilon (t)/dt \equiv \epsilon_t(t)= b/L^2 \sum_n s_n 
v_n(t)$ (the ``order parameter'') decays exponentially to zero 
for small $\sigma_\text{ext}$, while for larger $\sigma_\text{ext}$ a crossover to 
a steady state with a non-zero $\sigma_\text{ext}$-dependent $\epsilon_t$ can be 
observed. For an intermediate, critical value 
$\sigma_\text{ext}=\sigma_\text{c}$, $\epsilon_t$ decays as a power law of time,
$\epsilon_t \propto t^{-\theta}$, with $\theta \approx 1.0$. Thus, quenched disorder 
changes the large-scale dynamics of the system, as for a pure system one obtains the 
well-known 2$d$ Andrade law exponent $\theta \approx 2/3$ \cite{MIG-02}. The 
$\theta$-exponent has a value close to the mean-field depinning one (unity), but a 
glance at the insets of Fig. \ref{fig:2} reveals that the collective dislocation 
dynamics is not as simple as that would suggest. The cross-over time $t_\text{c}$ to 
the power-law relaxation regime decreases with increasing number $N_\text{s}$ of pinning 
sites, and the critical stress $\sigma_\text{c}$ obviously increases with $N_\text{s}$. 
Important is, however, that both the scalings are power-law -like. The behavior of 
$\sigma_\text{c} (N_\text{s})$ implies that the dislocations sample the quenched 
landscape collectively: the power-law relation is not linear in $N_\text{s}$ but scales 
with an exponent smaller than unity, close to 1/3. This is typical of random
manifolds, as is seen from Larkin length arguments in many cases \cite{MOR-04}.\\

\begin{figure}[!t]
\includegraphics[angle=0,width=8.5cm,clip]{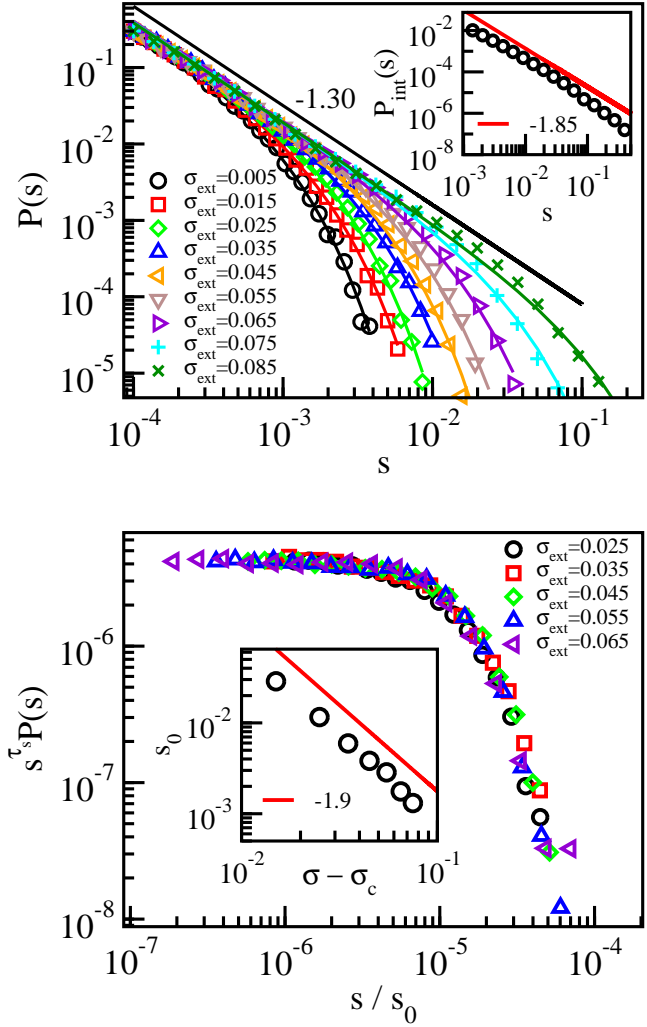}
\caption{\textbf{Critical scaling of the strain burst size distributions is depinning-like.} ({\bf a}) 
Distributions $P(s)$ of the slip avalanche sizes $s$ for various stress bins below the 
critical stress $\sigma_\text{c} \approx 0.09$, in a system with $N_\text{s} = 32000$ 
and $A = 0.1$, showing that the $\tau_s$-exponent has a value close to 1.30. The solid 
lines correspond to fits of equation (\ref{eq:ansatz}) with $f(x)=\exp{(-x)}$ to the data. 
The inset shows the corresponding stress-integrated distribution, with $\tau_{s,\text{int}}
\approx 1.85$. ({\bf b}) A data collapse of the $P(s)$ distibutions, with $\tau_s=1.3$ and 
$1/\sigma = 1.9$. The inset shows the cutoff avalanche size $s_\text{0}$ obtained from 
the fits shown in the top panel as a function of $\sigma_\text{c}-\sigma_\text{ext}$, 
confirming the value of $1/\sigma = 1.9$ used in the data collapse.}
\label{fig:3}
\end{figure}

{\bf \noindent Dislocation avalanches.}
Then we proceed to study deformation avalanches as a typical, experimental signature 
of criticality. Similarly to a micro-pillar compression experiment \cite{UCH-09,DIM-06} or 
recent numerical studies of the pure DDD model \cite{ISP-14}, we apply an adiabatic 
stress-ramp protocol such that individual, consequtive 
avalanches can be identified and analyzed. Thus we can follow the evolution of the 
deformation bursts all the way up to the yield stress. Starting as before from a relaxed 
configuration, $\sigma_\text{ext}$ is increased at a slow constant rate 
$\sigma_{\text{ext},t}$ (we consider $\sigma_{\text{ext},t}$-values ranging from 
$2.5\times10^{-7}$  to $2.5\times10^{-6}$), until the collective dislocation velocity 
$V(t) = \sum_n |v_n (t)|$ increases above a small threshold value $V_\text{th}=10^{-4}$. We define 
an avalanche as a continuous occurrence of $V(t) > V_\text{th}$, and keep $\sigma_\text{ext}$ 
constant until $V(t)$ falls again below $V_\text{th}$. The total strain increment 
$s = b/L^2 \sum s_n \Delta x_n$ accumulated during such an avalanche is taken to be 
the avalanche size $s$, and we also consider the statistics of the avalanche 
durations $T$. Once $V(t)<V_\text{th}$ and the avalanche has finished, the stress is 
again ramped up at a rate $\sigma_{\text{ext},t}$ until the next avalanche is triggered.

\begin{figure}[!t]
\includegraphics[angle=0,width=8.5cm,clip]{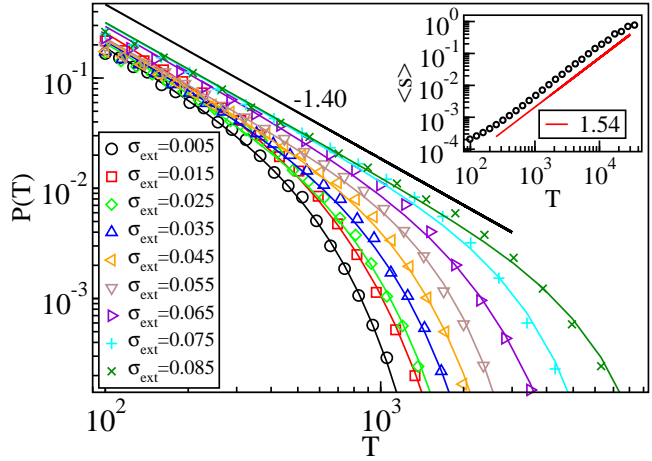}
\caption{\textbf{Criticality in durations of the deformation avalanches is in accordance 
with expectations from depinning phase transitions.} Distributions of the avalanche 
durations $P(T)$ for various stress bins below the critical stress $\sigma_c \approx 
0.09$ (with $N_s$ and $A$ as in Fig. \ref{fig:3}), showing that $\tau_T \approx 1.40$. 
The solid lines correspond to fits of the form $P(T) \propto T^{-\tau_T} \exp (-({T/T_0})^2)$. 
The inset shows the scaling of the average avalanche size $\langle s \rangle$ with the 
avalanche duration $T$, which follows $\langle s \rangle \propto T^{\gamma}$, with 
$\gamma \approx 1.54$. Notice that both $\tau_T$ and $\gamma$ have values that are 
clearly different from those of mean field depinning (i.e. 2 in both cases).}
\label{fig:4}
\end{figure}

The results obtained in the limit of a small threshold value $V_\text{th}=10^{-4}$ in Fig. 
\ref{fig:3} show that the data is described by the scaling
\begin{equation}
P(s,\sigma_\text{ext}) \propto s^{-\tau_s}f(s/s_0),
\label{eq:ansatz}
\end{equation}
where $\tau_s = 1.30 \pm 0.03$ and $s_0 \propto (\sigma_\text{c}-\sigma_\text{ext})^{-1/\sigma}$,
with $1/\sigma = 1.90 \pm 0.04$. Notice that this behaviour, while in agreement with the
standard depinning scaling picture, is fundamentally different from that observed in the
corresponding pure system, where $s_0$ is proportional to the exponential 
of the applied stress, $s_0 \propto \exp (\sigma_\text{ext}/\sigma_0)$, and the power law
exponent $\tau_s$ has a lower value $\tau_s \approx 1.0$ \cite{ISP-14}. We have checked that 
$\tau_s$, as well as the cutoff of the distribution of slip $sL^2$, are independent of the system 
size (see Supplementary Figs. S1 and S2).
The latter result is again in contrast to the pure system results, where the slip distribution
cutoff was found to exhibit a power law dependence on the number of dislocations or the
system size \cite{ISP-14}. Our estimates of $\tau_s$ and $1/\sigma$ are close but not equal to 
their mean-field depinning values ($3/2$ and 2, respectively) \cite{FIS-98}. The inset of 
the upper panel of Fig. \ref{fig:3} shows the stress-integrated distribution $P_\text{int}(s) 
\propto s^{-\tau_{s,\text{int}}}$, with $\tau_{s,\text{int}} = 1.85 \pm 0.10$. 
$\tau_{s,\text{int}}$ obeys within errorbars the scaling relation $\tau_{s,\text{int}} = 
\tau_s + \sigma$ \cite{DUR-06_2} and is also in reasonable agreement with the exponent 
value describing the distribution of dissipated energy during avalanches obtained from a 
minimal automaton model of 2$d$ crystal plasticity \cite{SAL-11}. It is worth noting that 
the range in which such critical scaling applies here (for the parameters $N_\text{s}$ 
and $A$ chosen in order to reduce any ``transient time'' as seen in the simulations 
with constant external stress, and to ensure a significant difference wrt. the 
disorder-free system) is very wide in external stress or the control parameter, in 
agreement with experiments. We have checked the robustness of our results by considering
three different values for $A$, all corresponding to
the ``pinning'' phase in Fig. \ref{fig:system}: all cases yield the same exponent 
characterizing the avalanche size distributions (see Supplementary Fig. S3).
Similar conclusions are reached when 
looking at the avalanche durations, $P(T,\sigma_\text{ext})$. Again (Fig. \ref{fig:4}), 
a wide scaling regime ensues. The data now indicates a $P(T) \propto T^{-\tau_T}g(T/T_0)$ -scaling 
with $\tau_T = 1.40 \pm 0.05$. The inset of Fig. \ref{fig:4} shows that the usual 
duration vs. size -relation of crackling noise holds, in that $\langle s(T) \rangle \propto 
T^{\gamma}$ with $\gamma = 1.54 \pm 0.05$ \cite{LAU-06}. Both these last exponents in 
particular, $\tau_T$ and $\gamma$, have values clearly different from their mean-field 
depinning counterparts (2 for both $\tau_T$ and $\gamma$).

\begin{figure}[!t]
\includegraphics[angle=0,width=8.5cm,clip]{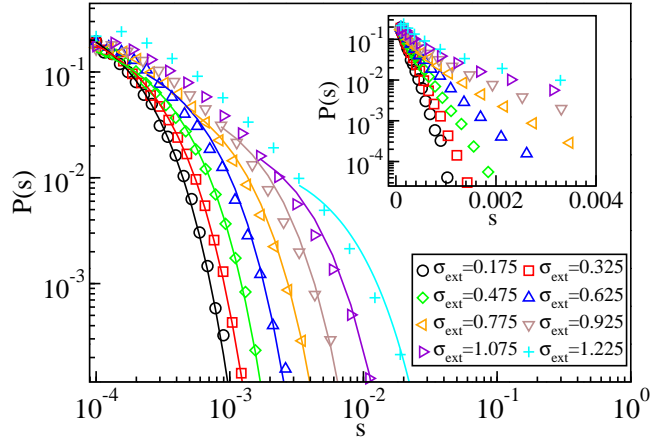}
\caption{\textbf{Very strong disorder quenches the critical behaviour.}
For $N_\text{s} = 32000$ and $A = 1.0$, the avalanche size distributions are no longer
power laws. Instead, a pure exponential function $P(s) \propto e^{-s/s_0}$ results
in a better fit. The inset shows the same distributions as in the main figure, but
with semi-logarithmic axes.}
\label{fig:5}
\end{figure}

Above we have shown that a depinning-like criticality can be established by fixing 
suitable, non-zero values for the disorder strength parameters $A$ and $N_s$. Obviously,
by lowering $A$ one approaches the disorder-free case of dislocation jamming. We do not 
look at the interesting issue how crossing the phase boundary looks like when moving from 
jamming to pinning (or vice versa). One would expect a kind of ``Larkin length'' to 
ensue, such that when a dislocation avalanche spans a large enough area to explore the 
random impurity landscape, it would show depinning-like characteristics instead of 
those related to jamming, consider again the insets of Fig.~\ref{fig:2}. It is a 
natural question to ask what happens if in the 
competition between long-range dislocation-dislocation interactions and the local effect 
of the pinning sites/solute atoms the latter starts to dominate. In Fig. \ref{fig:5} we 
show the outcome for $A = 1.0$, $N_s = 32000$, such that the forces experienced by the 
dislocations due to quenched pinning are much larger than those due to dislocation 
interactions. Now, all signs of power-law like avalanche activity are absent, and an 
exponential distribution is found, $P(s) \propto e^{-s/s_0}$, where 
$s_0 \propto e^{\sigma_\text{ext}/\sigma_0}$ and $\sigma_0 \approx 0.28$. Note that 
as expected, given the change of the $P(s)$, the avalanche sizes are now much more 
limited than in Fig. \ref{fig:3}, see also Supplemetary Movies 1 and 2, showing examples 
of the intermediate and strong disorder cases, respectively. A similar avalanche 
size-limiting effect due to strong pinning has been shown for superconducting vortex 
avalanches \cite{OLS-97}. This result confirms the third, strong-disorder phase 
stipulated in the phase diagram in Fig. \ref{fig:system}.

\vspace{0.5cm}
{\bf \large \noindent Discussion}

\noindent
In summary, we have studied the dynamics of 2$d$ dislocation assemblies in the presence of 
disorder. The model has a phase diagram (see Fig. \ref{fig:system}) that contains three phases.
As a special case one has the usual disorderless one \cite{ISP-14}, and then two where 
quenched disorder is important: A strong disorder one where collective dynamics does not 
exist, and another one with critical behavior typical of the depinning of elastic manifolds, 
and with a set of exponents different from the mean-field limit of this class of systems, despite
the long-range nature of the dislocation interactions. Our results leave fundamental questions 
about the phase diagram presented. We have argued that the mixing of long-range interactions 
and disorder leads to two new phases, one in which dislocation interactions are partly 
screened leading to depinning-like criticality with non-trivial exponents, and another where 
critical behavior is absent due to strong screening. The fundamental issues concern now the 
details of the phase diagram, the cross-overs from jamming to pinning and vice versa, and 
the precise location of the point where the jammed, pinned, and the flowing phase meet.

Which of the three phases is observed for a specific crystal should depend on the densities
of dislocations and pinning centers, and on the pinning strength induced by the latter: the 
fluctuating dynamics of a mobile dislocation is controlled by the spatial fluctuations 
of the forces of different origin experienced by it as it moves, i.e. those originating from 
dislocation interactions (a dynamic quantity), or from interactions with the static pinning 
centers. For instance, decreasing the dislocation density in a system with a fixed concentration
and strength of pinning centers will eventually lead to a situation where pinning forces
experienced by the dislocations dominate over those due to dislocation interactions. 
Another consequence of adding a quenched pinning field is given by the introduction of a 
microscopic disorder length-scale to the otherwise scale-free dislocation system, implying 
that the disordered dislocation system is not in the ``similitude regime'' \cite{ZAI-14}: 
an extreme manifestation of this is given by the lack of scale invariance of the strain 
bursts for very strong disorder.

Our study has been confined to the 2$d$ case for the basic reason that 3$d$ DDD models are 
numerically extremely challenging. There, in the absence of disorder, mean-field -like 
exponents have been claimed - although stress-resolved avalanche distributions were not 
considered in \cite{CSI-07} - and it follows naturally from our results that one expects 
to find the two screened phases also there with the introduction of point-like pinning 
centers. In 3$d$ systems, further complications may arise due to forest hardening: immobile
dislocations on inactive slip systems could have a similar effect as our quenched pinning 
centers, possibly leading to pinning-dominated dislocation dynamics even in the absense
of additional impurities such as solute atoms or precipitates. In BCC metals, also
sufficiently strong Peierls barriers may have a similar effect. Two-dimensional systems 
such as colloidal crystals \cite{PER-05} may provide relevant experimental systems to 
directly test our results. Note that a similar set of exponents to the one observed here 
for the intermediate disorder strength case was found very recently in a 2$d$ amorphous 
plasticity model \cite{BUD-13}, suggesting a possibility of a broad universality class 
of plastic deformation where microscopic details are irrelevant. 

A most important practical conclusion is that the microstructure of materials with 
dislocation activity may induce discrete qualitative changes in the bursty dynamics: 
jamming or pinning. The depinning phase should give rise to usual phenomena such as 
thermally assisted creep \cite{CHA-00} and glassy relaxation \cite{CUG-96} which relate 
to the critical exponents of the transition, and where the spatial correlations (point- 
or line-like and so forth) of the disorder are relevant. 
An obvious further generalization is to time-dependent disorder such as diffusing solute 
fields \cite{WAN-00}, where phenomena such as the Portevin-Le Chatelier effect should ensue 
\cite{POR-23, ZAI-97}.

\vspace{0.5cm}
{\bf \large \noindent Methods}

{\bf \noindent 2D DDD model with pinning.}
The $2d$ DDD model we study is a development to other models studied in the
literature \cite{GIE-95,MIG-08}, with the addition of a random arrangement of $N_\text{s}$
quenched pinning centers. It represents a cross section ($xy$ plane) of a single
crystal with a single slip geometry and straight parallel edge dislocations
along the $z$ axis. The $N_\text{d}$ edge dislocations glide along directions parallel
to their Burgers vectors ${\bf b}=\pm b {\bf u}_x$, where $b$ is the magnitude and 
${\bf u}_x$ is the unit vector along the $x$ axis. Equal numbers of
dislocations with positive and negative Burgers vectors are assumed, and
dislocation climb is not considered: The latter is a good approximation in low
temperatures \cite{KER-12,BAK-07}. The dislocations
interact with each other through their long-range stress fields,
$\sigma_\text{d}({\bf r}) = Db x(x^2 - y^2)/(x^2+y^2)^2$,
where $D=\mu/2\pi(1-\nu)$, with $\mu$ the shear modulus and $\nu$
the Poisson ratio of the material. In addition, we consider a random
arrangement of $N_\text{s}$ immobile solute atoms interacting with the
dislocations via short-range interactions. To this end, we use the regularized
interaction energy derived from non-local elasticity \cite{WAN-90}, expressed in
polar coordinates for the $n$th dislocation with sign $s_n=\pm 1$ as
\begin{equation}
U_\text{NL}^n=-\frac{(1+\nu)\mu s_n b \Delta V}{3\pi (1-\nu)}\frac{\sin \theta}{r}
\left[ 1-\exp \left( -\frac{k^2}{a^2}r^2 \right) \right], \label{eq:s}
\end{equation}
where $\Delta V$ is the misfit area, $k=1.65$, and $a$ is the atomic
distance \cite{WAN-90}. This regularized form of the interaction energy removes
the singularity at $r=0$. Other short-range pinning potentials should
lead to similar results: we have checked that this is true for Gaussian
pinning centers with $U=-Ae^{-r^2/R^2}$ (see Supplementary Fig. S4).
The corresponding interaction force
acting on the dislocations due to a solute atom is then given by 
${\bf F}_\text{ds}=-\nabla U_\text{NL}$.

Thus, the overdamped equations of motion of the dislocations read
\begin{eqnarray}
\frac{v_n}{\chi_d b} & = & s_n b \left[ \sum_{m \neq n}^{N_d} \sigma_\text{d}^m({\bf r}_{nm}) +
\sigma_\text{ext} \right] \\ \nonumber
& & -\frac{1}{b}\sum_i^{N_s} (\nabla U_\text{NL}^n)\cdot {\bf u}_x \label{eq:d}
\end{eqnarray}
with $v_n$ the velocity and $s_\text{n}$ the sign of the
$n$th dislocation, $\chi_\text{d}$ the dislocation mobility (implicitly
including effects due to thermal fluctuations), and $\sigma_\text{ext}$ is
external stress. The dislocation-solute force decays with distance
as $1/r^2$ (while the dislocation-dislocation force $\sim 1/r$), and we
introduce a cut-off distance $r_\text{cutoff}=15b$ (corresponding typically to
two times the average dislocation-dislocation distance) beyond which the
dislocation-solute interaction is set to zero. At the cut-off distance, the
dislocation-solute interaction is several orders of magnitude
smaller than typical dislocation-dislocation interactions, and thus
has a negligible effect on dislocation dynamics. The equations of motion are integrated with an
adaptive step size fifth order Runge-Kutta algorithm, by measuring lengths
in units of $b$, times in units of $1/(\chi_\text{d} Db)$, and stresses in units
of $D$, and by imposing periodic boundary conditions in the $x$
direction. Two dislocations of opposite sign with a mutual distance smaller
than $b$ are removed from the system, to include a mechanism for dislocation
annihilation in the model.

The simulations are started from a random initial configuration
of $N_\text{d}=1600$ dislocations within a square cell of linear size
$L=200b$. These initial states are first relaxed with $\sigma_\text{ext} = 0$,
to reach metastable dislocation arrangements; Fig. \ref{fig:system} shows a
local detail of such a system. After the annihilations during
the relaxation, $N_\text{d} \approx 800-900$ dislocations remain. Then, an
external stress is turned on, and the evolution of the system is
monitored by measuring the time dependence of various quantities, such
as the strain rate, $d\epsilon (t)/dt \equiv \epsilon_t(t)= b/L^2 \sum_n s_n v_n(t)$.
In the simulations, we consider the effect of varying both the dislocation-solute
interaction strength $A=(1+\nu)\mu b \Delta V/3 \pi (1-\nu)$ and the solute
density $\rho_s = N_\text{s}/L^2$. In the absence of correlations, $A\rho_s$ measures
the relative strength of disorder, compared to the dislocation interactions.
The range of values considered for the disorder is $A=0.05-1.0$ and $N_\text{s}=500-32000$,
giving $A\rho_s \approx 10^{-3}-1.0$. The results are averaged over
a large number of realizations for each set of parameters, ranging from 500 to
6000.

\vspace{0.5cm}
\noindent{\bf Supplementary Information} is linked to the online version
of the paper

\vspace{0.5cm}
\noindent{\bf Acknowledgements}\\
P. Isp\'anovity, M. Zaiser, S. Zapperi, and I. Groma are
thanked for discussions. We acknowledge the financial support of the Academy
of Finland through an Academy Research Fellowship (L.L., project no. 268302) and 
the Centres of Excellence Program (project no. 251748). The numerical simulations 
presented above were performed using computer resources within the Aalto University 
School of Science ``Science-IT'' project.

\vspace{0.5cm}
\noindent{\bf Author contributions}\\
M.O., L.L. and M.J.A designed the study. M.O. performed the numerical modelling
and data analysis. L.L. wrote the first draft of the manuscript.
All authors contributed to improve the manuscript.

\vspace{0.5cm}
\noindent{\bf Competing financial interests}\\
The authors declare no competing financial interests.

\end{document}